\documentclass[pdflatex,sn-basic]{sn-jnl}

\usepackage{graphicx}%
\usepackage{multirow}%
\usepackage{amsmath,amssymb,amsfonts}%
\usepackage{amsthm}%
\usepackage{mathrsfs}%
\usepackage[title]{appendix}%
\usepackage{lmodern}
\usepackage{xcolor}%
\usepackage{textcomp}%
\usepackage{manyfoot}%
\usepackage{booktabs}%
\usepackage{algorithm}%
\usepackage{algorithmicx}%
\usepackage{algpseudocode}%
\usepackage{listings}%

\theoremstyle{thmstyleone}%
\theoremstyle{thmstyletwo}%

\theoremstyle{thmstylethree}%

\begin{document}

\title[Article Title]{Universal Matrix Multiplication on Quantum Computer}


\author[1]{Jiaqi Yao}

\author[2]{Tianjian Huang}

\author[3]{Zipeng Cai}

\author*[4]{Ding Liu}\email{liuding@tiangong.edu.cn}

\affil{\orgdiv{School of Computer Science and Technology}, \orgname{Tiangong University}, \orgaddress{ \city{Tianjin}, \postcode{300387}, \country{China}}}




\abstract{As the most central and computationally intensive component of deep neural networks, the execution efficiency of matrix multiplication directly determines the training and inference performance of models. Harnessing the parallel processing capabilities afforded by quantum superposition and entanglement to reshape matrix multiplication implementations has become a promising entry point for optimising underlying quantum arithmetic logic and improving the operational efficiency of quantum circuits. This paper proposes a universal quantum matrix multiplication (QMM) framework designed to achieve substantial computational acceleration through an optimised quantum arithmetic logic unit. To circumvent the limitations of multi-register and multi-control gates in conventional quantum arithmetic circuits, we encode classical data directly into parameterised \(R_z\) rotation gates using the quantum Fourier transform (QFT), thereby reducing the base gate complexity of the quantum adder to \(O(n)\). In addition, by adopting the column-wise multiplication principle from classical arithmetic, we optimize the gate complexity of the quantum multiplier to \(O(n^2)\). We further extend this approach to a quantum version of the Strassen algorithm, and experimentally quantify the trade-off between reduced multiplication time and increased overhead in addition resources. This work establishes a reliable technical pathway for constructing general-purpose quantum matrix operations, with the potential to unlock substantial computational power for training modern machine learning models.}




\maketitle

\section{Introduction}\label{sec1}

In the era of artificial intelligence driven by big data, the scale of deep learning models is growing at an exponential rate. Matrix multiplication, as a core operation in scientific computing and deep learning, has become a critical factor limiting overall system performance. However, as Moore’s law slows, classical computers face severe computational bottlenecks when handling trillion-parameter models. Quantum computing, with its unique parallelism and superposition properties, offers a new paradigm for overcoming this bottleneck.

Classical matrix multiplication has been steadily pushing the theoretical limit for decades. In 1969, Strassen \cite{strassen1969gaussian} first reduced the time complexity of matrix multiplication to \(O(n^{\log 7})\). In 1987, Winograd and Coppersmith \cite{coppersmith1987matrix}  extended the Strassen algorithm \cite{pan1978strassen,coppersmith1982asymptotic,strassen1987relative}, achieving a complexity of \(O(n^{2.376})\). Subsequently, researchers devoted considerable effort to the continuous optimization of these classical frameworks \cite{stothers2010complexity,williams2012multiplying,le2014powers,alman2021refined}, leading to frequent updates of the complexity upper bound. In 2022, DeepMind introduced AlphaTensor \cite{fawzi2022discovering}, a model that leverages deep reinforcement learning to automatically search for more efficient matrix multiplication algorithms. In a more recent study, reference \cite{duan2023faster} improved the notation-block technique within the laser method, further optimizing the matrix multiplication complexity to \(O(n^{2.371866})\).

Meanwhile, research on matrix multiplication in the quantum computing domain has also achieved continuous breakthroughs. Building on Le Gall's review of quantum algorithms in the semiring setting \cite{le12quantum}, initial efforts focused on optimizing Boolean matrix multiplication. Buhrman and Š palek \cite{buhrman2004quantum} used quantum random walk techniques to reduce the complexity to \(O(n^{3/2}\sqrt{l})\), where \(l\) is the number of non-zero entries in the product \(AB\) and \(n\) is the dimension of the input matrices. Subsequently, schemes based on the Lingas algorithm \cite{le2012improved,lingas2011fast} and the graph collision framework \cite{jeffery2016improving,le2014quantum} were proposed, further compressing query and time complexities. Beyond Boolean operations, quantum advantages are particularly pronounced in specialized matrix multiplications; for example, reference \cite{le2014quantum} improved the complexity of max–min matrix multiplication and distance matrix multiplication to \(\tilde{O}(n^{2.473})\) and \(\tilde{O}(2^{0.64l}n^{2.46})\), respectively, both surpassing the then best classical algorithms \cite{duan2009fast,vassilevska2006finding,yuster2009efficient}. In addition, for the estimation of general matrix products, researchers have explored diverse quantum implementation routes based on the swap test \cite{buhrman2001quantum}, singular value transformation (SVE) \cite{kerenidis2016quantum}, and the HHL algorithm \cite{lloyd2010quantum} among others \cite{shao2018quantum}, providing rich theoretical support for quantum-accelerated large-scale numerical computation.

Unlike previous studies that focused on specific structures or Boolean matrices, here we propose a general-purpose quantum matrix multiplication framework based on quantum arithmetic operations, capable of handling arbitrary integer data. The core of this scheme lies in the design of efficient quantum adders and multipliers. In 2000, Draper \cite{draper2000addition} proposed a quantum Fourier transform (QFT)-based adder to reduce the number of qubits required for addition. Subsequently, using QFT to optimize arithmetic logic units (ALUs) and achieve efficient quantum addition and multiplication has become a research focus \cite{beauregard2002circuit,pavlidis2012fast,pavlidis2021quantum}. Following the QFT adder, Ruiz-Perez and Garcia-Escartin \cite{ruiz2017quantum} in 2017 further constructed multiplication logic in Fourier space using double-controlled phase gates, thereby enabling effective quantization of multiplication operations.

This paper proposes a general-purpose quantum matrix multiplication framework that uses QFT to encode classical data into Fourier basis states and draws on classical arithmetic logic to realise quantised matrix operations. By encoding data directly into \(R_z\) rotation gates and adopting the column-wise multiplication design principle, we substantially reduce circuit depth and hardware overhead. Theoretical analysis shows that our scheme reduces the elementary gate complexity of the adder and multiplier to \(O(n)\) and \(O(n^2)\), respectively. Having validated the feasibility and acceleration performance of the underlying arithmetic operators, we further explore a quantised implementation of the Strassen algorithm and comparatively analyse its advantages and disadvantages over classical algorithms in complex operations.

\begin{table*}
\caption{Resource consumption for a single addition and multiplication operation by the optimized adder and multiplier. ($n$ represent the number of qubits used to encode the integers.)}\label{tab1}
\begin{tabular*}{\textwidth}{@{\extracolsep\fill}lcccccc}
\toprule%
& \multicolumn{3}{@{}c@{}}{adder} & \multicolumn{3}{@{}c@{}}{multiplier} \\\cmidrule{2-4}\cmidrule{5-7}%
& original & optimized  & 
classical  & original  & optimized  & classical \\
\midrule
qubits  & $2n+1$  & $n+1$ & - & $4n$  &  $3n$  & -\\
gates  & $\frac{n^2+3n}{2}$ & $n+1$  & $5n-3$ & $2n^3$ & $\frac{3n^2+n}{2}$ & $6n^2$\\
\toprule
\end{tabular*}
\end{table*}

\section{Method}
\subsection{Quantum Fourier Transform}
The Quantum Fourier Transform (QFT) is a powerful quantum algorithm designed to perform Fourier transformations on quantum amplitudes \cite{ruiz2017quantum,nielsen2001quantum,weinstein2001implementation}. While QFT doesn't accelerate classical tasks like Fourier transformations on classical data, it serves the purpose of encoding data effectively.

The Quantum Fourier Transform is a linear operator defined on a set of standard orthogonal bases $\left| 0 \right>,\cdots ,\left| N-1 \right>$. Its action on the basis states is as follows:
\begin{equation}
{\left| j \right> \rightarrow \frac{1}{\sqrt{N}}\sum_{k=0}^{N-1}{e^{\dfrac{2\pi ijk}{N}}\left| k \right>}}
\label{eq-QFT}
\end{equation}
equivalently, the action on any state can be expressed as:
\begin{equation}
{\sum_{j=0}^{N-1}{x_j\left| j \right>}\rightarrow \sum_{k=0}^{N-1}{y_k\left| k \right>}}
\label{eq-QFTstate}
\end{equation}
whereby, the amplitude $y_k$ represents the values after performing the discrete Fourier transform on the amplitudes.

Let $N=2^n$, where $n$ represents the number of qubits in a quantum computer. Considering $N$ orthogonal normalized states $\left| 0 \right> ,\cdots ,\left| 2^n-1 \right> $ as the computational basis states in the Hilbert space, we express the state $\left| j \right> $ in binary form as $j=j_1j_2\cdots j_n$, which is equivalent to $j=j_12^{n-1}+j_22^{n-2}+\cdots +j_n2^0$. For simplicity, we use $0.j_lj_{l+1}\cdots j_n$ to denote the binary fraction $\dfrac{j_l}{2}+\dfrac{j_{l+1}}{2^2}+\cdots +\dfrac{j_n}{2^{n-l+1}}$. Through algebraic operations, we derive the product form of the Quantum Fourier Transform:
\begin{equation}
\begin{aligned}
&\left| j_1,\cdots ,j_n \right> \\&\xrightarrow{QFT}\frac{1}{\sqrt{N}}\sum_{k=0}^{N-1}{e^{\dfrac{2\pi ijk}{N}}\left| k \right>}
\\&=\frac{1}{\sqrt{2^n}}\sum_{k_1=0}^1{\cdots \sum_{k_n=0}^1{e^{\dfrac{2\pi ij\left( k_12^{n-1}+\cdots +k_n2^0 \right)}{2^n}}}\left| k_1\cdots k_n \right>}
\\&=\frac{1}{\sqrt{2^n}}\sum_{k_1=0}^1{\cdots \sum_{k_n=0}^1{e^{2\pi ij\left( k_12^{-1}+\cdots +k_n2^{-n} \right)}}\left| k_1\cdots k_n \right>}
\\&=\bigotimes_{l=1}^n{\frac{\left| 0 \right> +e^{2\pi i\left( j2^{-l} \right)}\left| 1 \right>}{\sqrt{2}}}
\\&=\frac{\left| 0 \right> +e^{2\pi i0.j_n}\left| 1 \right>}{\sqrt{2}}\otimes \cdots \otimes \frac{\left| 0 \right> +e^{2\pi i0.j_1j_2\cdots j_n}\left| 1 \right>}{\sqrt{2}}
\label{eq-QFTs}
\end{aligned}
\end{equation}

\begin{figure*}
	\centering
	\includegraphics[width=1.0\textwidth
    ]{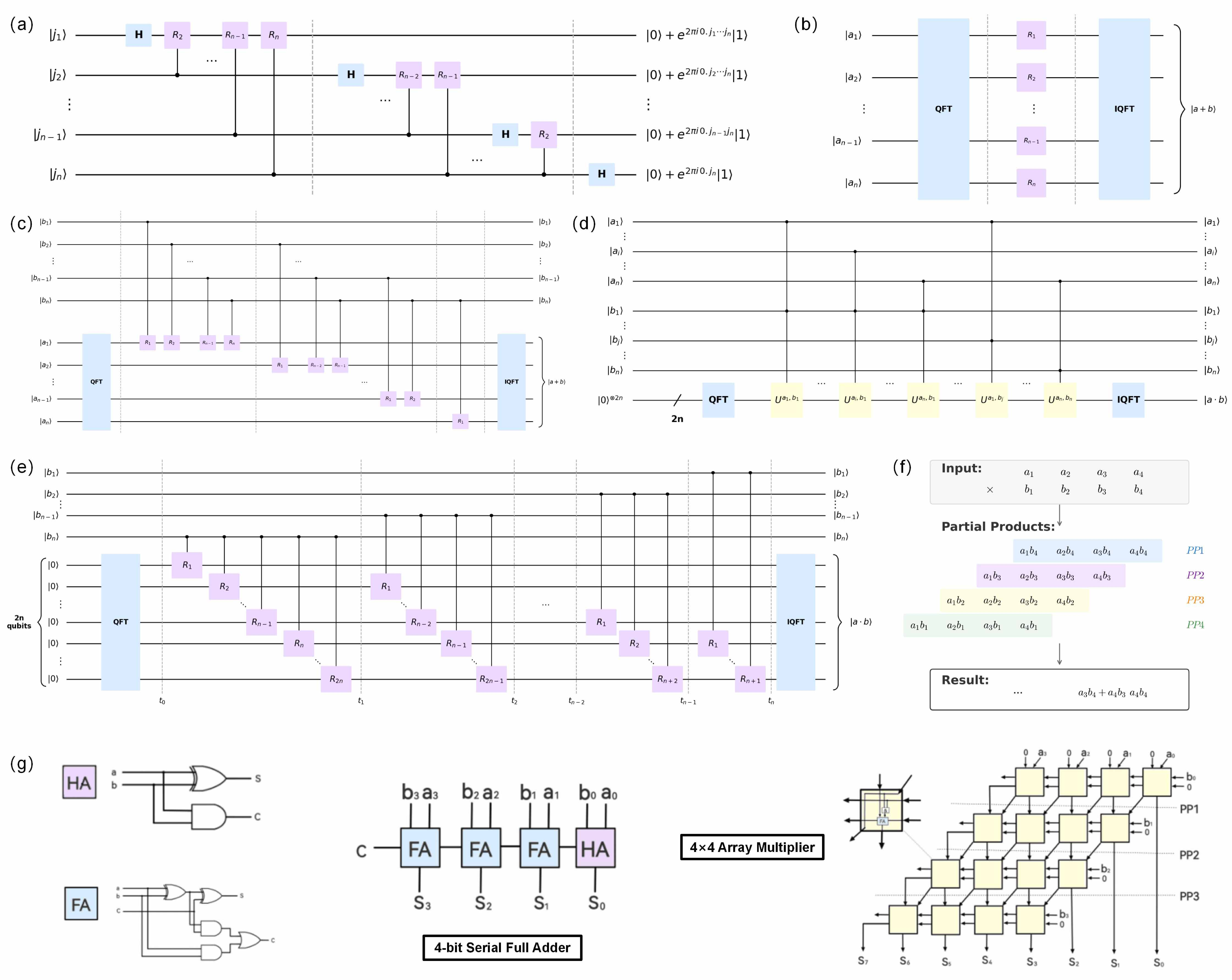}
	\caption{Circuit structures and logical principles of quantum arithmetic units; (a) Circuit for the Quantum Fourier Transform; (b) Circuit for the optimized quantum adder; (c) Circuit for the original QFT-based quantum adder; (d) Circuit for the original quantum multiplier; (e) Circuit for the optimized quantum multiplier; (f) Schematic diagram of classical column-wise multiplication logic; (g) Implementations of 4-bit serial full adder and $4\times4$ array multiplier.}
	\label{fig:Fig1_cir}
\end{figure*}

The quantum circuit for the Quantum Fourier Transform is depicted in Fig.\ref{fig:Fig1_cir}(a), incorporating controlled-phase gates and Hadamard gates.
\begin{equation}
R_k\equiv \left[ \begin{matrix}
	1&		0\\
	0&		e^{\dfrac{2\pi i}{2^k}}\\
\end{matrix} \right] \ \ \ and\ \ \ H=\frac{1}{\sqrt{2}}\left[ \begin{matrix}
	1&		1\\
	1&		-1\\
\end{matrix} \right] 
\label{eq-RxHgate}
\end{equation}
When the input state is $\left| j_1,\cdots ,j_n \right> 
$, the transformation process of the first qubit unfolds as follows:
\begin{equation}
\begin{aligned}
\left| j_1 \right> 
&\rightarrow \frac{1}{\sqrt{2}}\left( \left| 0 \right> +e^{2\pi i0.j_1}\left| 1 \right> \right) \ \ \ \ \ \ \ \ \ Hadamard\ gate
\\&\rightarrow \frac{1}{\sqrt{2}}\left( \left| 0 \right> +e^{2\pi i0.j_1j_2}\left| 1 \right> \right) \ \ \ \ \ \ \ R_2\ gate
\\&\ \ \ \ \ \ \ \ \ \ \ \ \ \ \ \ \vdots \ \ \ \ \ \ \ \ \ \ \ \ \ \ \ \ \ \ \ \ \ \ \ \ \ \ \ \ \ \ \vdots 
\\&\rightarrow \frac{1}{\sqrt{2}}\left( \left| 0 \right> +e^{2\pi i0.j_1j_2\cdots j_n}\left| 1 \right> \right) \,\,\,\,\,\,\,R_n\,\,gate
\label{eq-QFTqubit}
\end{aligned}
\end{equation}

From a mathematical perspective, the Fourier transform can also be understood as a change of basis. In other words, through the QFT transformation, numbers can be represented using Fourier bases. In this scenario, states are represented through qubits on the XY plane of the Bloch sphere. According to the Eq.\ref{eq-QFT}, assuming we are using n qubits and we wish to represent the number m using the Fourier basis, the phase of the j-th qubit would be $\alpha _j=\frac{2\pi m}{2^j}$.

\subsection{Quantum Adder}
\subsubsection{The Original Quantum Adder}
The reference \cite{draper2000addition} proposes that quantum addition is performed through a series of mutually commutable conditional rotations, which is very similar to the Quantum Fourier Transform. Therefore, quantum addition can be realized using the QFT. The specific quantum circuit is illustrated in Fig.\ref{fig:Fig1_cir}(c). Specifically, the first operand is first encoded from the computational basis to the Fourier basis, then rotations are performed on the encoded qubits to carry out addition in the Fourier basis. Finally, the sum is transformed from the Fourier basis back to the computational basis using the Quantum Fourier Inverse Transform. Integrated with the QFT circuit mentioned above, the transformation process of the first qubit is as follows:
\begin{equation}
\begin{aligned}
& \ \ \ \ \ \ \frac{1}{\sqrt{2}}\left( \left| 0 \right> +e^{2\pi i0.j_1j_2\cdots j_n}\left| 1 \right> \right) 
\\&\rightarrow \ \frac{1}{\sqrt{2}}\left( \left| 0 \right> +e^{2\pi i\left( 0.j_1j_2\cdots j_n+0.p_1 \right)}\left| 1 \right> \right) \,\,\,\ \ \ \ \ \ \,\,\ \ CR_1
\\&\rightarrow \,\,\frac{1}{\sqrt{2}}\left( \left| 0 \right> +e^{2\pi i\left( 0.j_1j_2\cdots j_n+0.p_1p_2 \right)}\left| 1 \right> \right) \,\,\,\,\ \,\,\,\,\,\,\,CR_2
\\&\ \ \ \ \ \ \ \ \ \ \ \ \ \ \ \ \ \ \ \ \ \ \ \ \ \ \ \,\,\,\,\,\,\,\,\,\,\vdots \,\,\,\,\,\,\,\,\,\,\,\,\,\,\ \ \ \ \ \ \ \ \ \ \ \ \ \ \ \ \ \,\,\,\,\,\,\,\,\,\,\,\,\,\,\,\,\vdots 
\\&\rightarrow \,\,\frac{1}{\sqrt{2}}\left( \left| 0 \right> +e^{2\pi i\left( 0.j_1j_2\cdots j_n+0.p_1p_{2\cdots}p_n \right)}\left| 1 \right> \right)\,\,\,\,\ CR_n
\label{eq-QFTfirstadd}
\end{aligned}
\end{equation}

\subsubsection{The Optimized Quantum Adder}
In the preceding algorithm, data extraction from an additional register precedes addition calculations. However, it's noted that during many intermediate computational steps, direct manipulation of intermediate variables on rotation gates is feasible \cite{ruiz2017quantum}. As previously discussed, utilizing $n$ qubits $(n>m)$ in the Fourier basis enables the representation of any number $m$, with the j-th qubit corresponding to a phase of $\frac{2\pi m}{2^j}$ . Given the phase-aligned states encoded by QFT, significant flexibility is afforded during arithmetic operations. It is posited that with the assumption of a singular addition operation and sufficient qubits (qubits$=n+1$), direct phase manipulation of the second addend post the Quantum Fourier Transform of the first addend is viable. The specific quantum circuit is delineated in Fig.\ref{fig:Fig1_cir}(b)

However, in practical computational scenarios, multiple data additions may be encountered. To mitigate the need for repetitive QFT encoding of data and to conserve quantum gates, it is feasible to extend the adder for a single addition operation. Following the Quantum Fourier Transform, multiple phase shifts can be executed, culminating in the utilization of the inverse Quantum Fourier Transform to output the result.

\begin{figure*}
	\centering
	\includegraphics[width=0.7\textwidth]{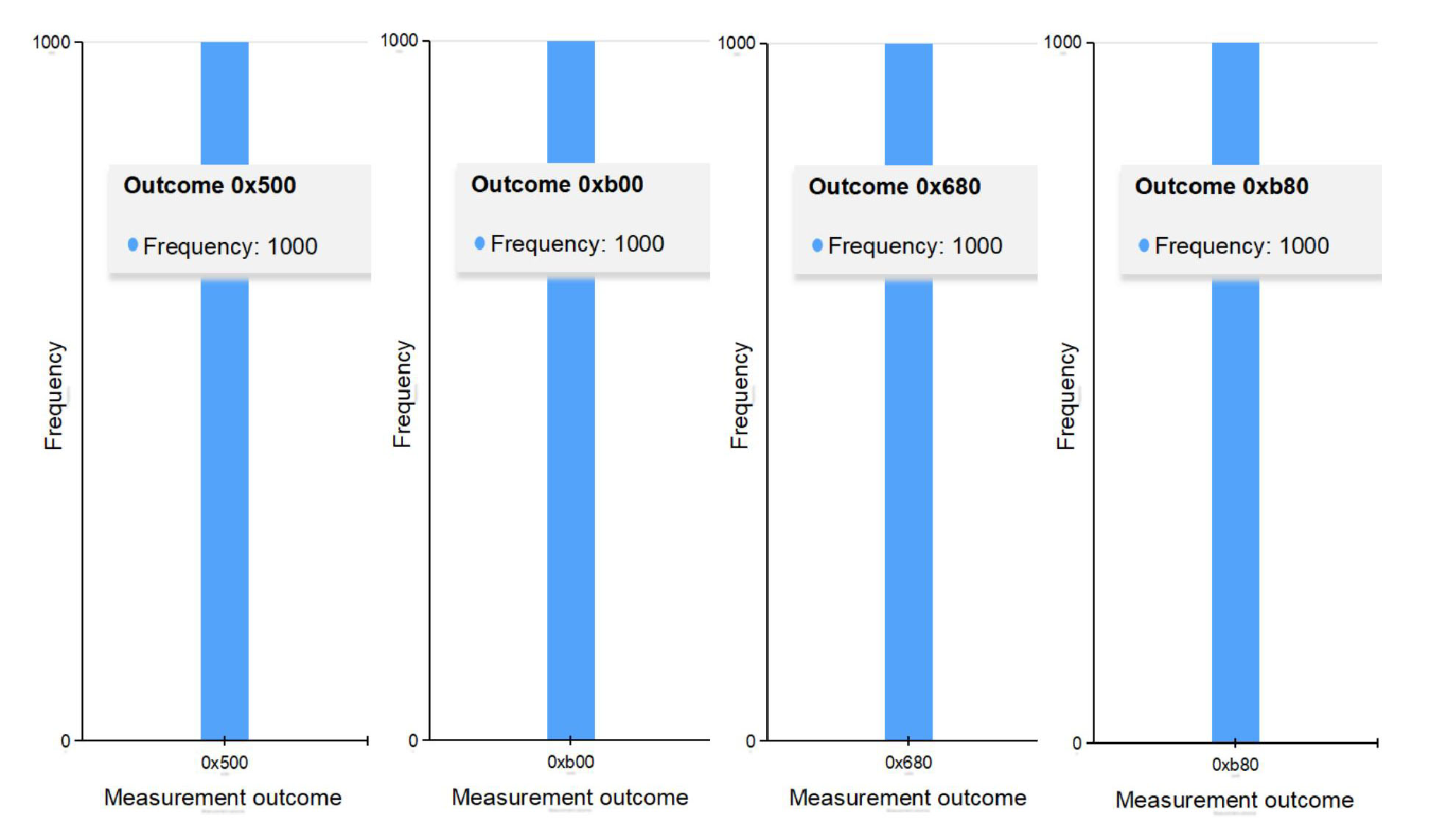}
	\caption{Measurement results of the quantum simulation for the matrix product $C=AB$, where $A=\left[\begin{matrix}1&2\\3&4\\ \end{matrix} \right]$ and $B=\left[\begin{matrix}2&3\\4&5\\ \end{matrix} \right]$. (The results are in hexadecimal and represented in reverse order.)}
	\label{fig:Fig2_result}
\end{figure*}

\subsection{Quantum Multiplier}
\subsubsection{The Original Quantum Multiplier}
The implementation principles of the multiplier and adder are consistent, as both rely on QFT. To better describe the working principle of the multiplier, we represent the two multiplicands in binary as $a=a_12^{n-1}+a_22^{n-2}+\cdots +a_n2^0$ and $b=b_12^{n-1}+b_22^{n-2}+\cdots +b_n2^0$. From this, we can derive the formula:
\begin{equation}
\begin{aligned}
c&=a\cdot b
\\& =\left( a_12^{n-1}+\cdots +a_n2^0 \right) \left( b_12^{n-1}+\cdots +b_n2^0 \right) 
\\& =\sum_{i=1}^n{a_i2^{n-i}}\cdot \sum_{j=0}^n{b_j2^{n-j}}
\\&=\sum_{i=1}^n{\sum_{j=1}^n{a_ib_j2^{2n-i-j}}}
\label{eq-mul}
\end{aligned}
\end{equation}
where n represents the number of qubits used to encode the integers a and b. When $a_i=1$ and $b_j=1$, the coefficient of the binary result 
$2^{2n-i-j}$ is 1 \cite{ruiz2017quantum}. During the experimental setup, both multiplicands $a $ and $b$ need to be encoded as binary data. The result $c$ is encoded into the Fourier basis using the QFT, followed by operations on controlled rotation gates. The calculation result is obtained through the Quantum Fourier Inverse Transform. The implementation of the quantum multiplier requires $4n$ qubits. The specific quantum circuit is depicted as shown in Fig.\ref{fig:Fig1_cir}(d).

\subsubsection{The Optimized Quantum Multiplier}
The optimized quantum multiplier leverages the concept of column multiplication (Fig.\ref{fig:Fig1_cir}(f)). Similarly, the two multiplicands are represented in binary as $a=a_12^{n-1}+a_22^{n-2}+\cdots +a_n2^0$ and $b=b_12^{n-1}+b_22^{n-2}+\cdots +b_n2^0$. The multiplication result of the multiplier is achieved through shifting-addition of partial products. Taking the partial product PP1 as an example, 
\begin{equation}
\begin{aligned}
PP1=a\cdot b_n=\left( a_12^{n-1}+\cdots +a_n2^0 \right) b_n
\end{aligned}
\end{equation}    
if $b_n=1$, then $PP1=a$, and if $b_n=0$, then $PP1=0$, and a single controlled rotation gate precisely accomplishes this usage. 
In the optimized adders, it is understood that with an ample supply of qubits, encoding any value becomes feasible through complete qubit rotations within the output register. Hence, we can directly treat the multiplicand as an intermediate variable and manipulate it within rotation gates. Consequently, there's no need to compute the product of each pair of single-bit positions, as in classical multipliers. Instead, the sum of partial products can be directly computed using single-controlled rotation gates.

The quantum circuit of the optimized multiplier is depicted in Fig.\ref{fig:Fig1_cir}(e).  Performing the Quantum Fourier Transform on the number a using n qubits yields (eq.\ref{eq-QFTs}):
\begin{equation}
\begin{aligned}
QFT\left| a \right> & =\frac{1}{\sqrt{2^n}}\sum_{k=0}^{2^n-1}{e^{\frac{2\pi iak}{2^n}}\left| k \right>}
\\&=\bigotimes_{l=1}^n{\frac{\left| 0 \right> +e^{\frac{2\pi ia}{2^l}}\left| 1 \right>}{\sqrt{2}}}
\end{aligned}
\end{equation} 
In the quantum circuit, we pass the multiplicand as an intermediate variable to the rotation gates, ensuring that the corresponding phase of the j-th qubit is $\alpha _j=\frac{2\pi a}{2^j}$. The derivation formula for the output qubit at each moment is:
\begin{equation}
\begin{aligned}
& \xrightarrow{t_0}\bigotimes_{l=1}^{2n}{\left| 0 \right>}
\\& \xrightarrow{t_1}\frac{\left| 0 \right> +e^{2\pi ia\cdot \frac{b_n}{2^1}}\left| 1 \right>}{\sqrt{2}}\otimes \frac{\left| 0 \right> +e^{2\pi ia\cdot \frac{b_n}{2^2}}\left| 1 \right>}{\sqrt{2}} \otimes \cdots \otimes
\frac{\left| 0 \right> +e^{2\pi ia\cdot \frac{b_n}{2^{n-1}}}\left| 1 \right>}{\sqrt{2}}
\\& \qquad \otimes \cdots \otimes \frac{\left| 0 \right> +e^{2\pi ia\cdot \frac{b_n}{2^{2n}}}\left| 1 \right>}{\sqrt{2}}
\\& \xrightarrow{t_2}\frac{\left| 0 \right> +e^{2\pi ia\cdot \frac{b_n}{2^1}}\left| 1 \right>}{\sqrt{2}}\otimes \frac{\left| 0 \right> +e^{2\pi ia\cdot \left( \frac{b_n}{2^2}+\frac{b_{n-1}}{2^1} \right)}\left| 1 \right>}{\sqrt{2}}
\otimes \cdots \otimes  
\\& \qquad \frac{\left| 0 \right> +e^{2\pi ia\cdot \left( \frac{b_n}{2^{n-1}}+\frac{b_{n-1}}{2^{n-2}} \right)}\left| 1 \right>}{\sqrt{2}}\otimes \cdots \otimes
\frac{\left| 0 \right> +e^{2\pi ia\cdot \left( \frac{b_n}{2^{2n}}+\frac{b_{n-1}}{2^{2n-1}} \right)}\left| 1 \right>}{\sqrt{2}}
\\& \ \ =\frac{\left| 0 \right> +e^{2\pi ia\cdot \frac{b_n}{2^1}}\left| 1 \right>}{\sqrt{2}}\otimes \frac{\left| 0 \right> +e^{2\pi ia\cdot \left( \frac{b_n+2b_{n-1}}{2^2} \right)}\left| 1 \right>}{\sqrt{2}}
\otimes \cdots \otimes 
\\& \qquad \frac{\left| 0 \right> +e^{2\pi ia\cdot \left( \frac{b_n+2b_{n-1}}{2^{n-1}} \right)}\left| 1 \right>}{\sqrt{2}}
\otimes \cdots \otimes
\frac{\left| 0 \right> +e^{2\pi ia\cdot \left( \frac{b_n+2b_{n-1}}{2^{2n}} \right)}\left| 1 \right>}{\sqrt{2}}
\\& \qquad \qquad \qquad \qquad \qquad \qquad \vdots
\end{aligned}
\end{equation}

\begin{equation}
\begin{aligned}
\\& \xrightarrow{t_{n-1}}\frac{\left| 0 \right> +e^{2\pi ia\cdot \frac{b_n}{2^1}}\left| 1 \right>}{\sqrt{2}}\otimes \frac{\left| 0 \right> +e^{2\pi ia\cdot \left( \frac{b_n}{2^2}+\frac{b_{n-1}}{2^1} \right)}\left| 1 \right>}{\sqrt{2}}
\otimes \cdots \otimes  
\\& \qquad \frac{\left| 0 \right> +e^{2\pi ia\cdot \left( \frac{b_n}{2^{n-1}}+\frac{b_{n-1}}{2^{n-2}}+\cdots +\frac{b_2}{2^1} \right)}\left| 1 \right>}{\sqrt{2}}\otimes \cdots \otimes 
\\& \qquad \frac{\left| 0 \right> +e^{2\pi ia\cdot \left( \left( \frac{b_n}{2^{2n}}+\frac{b_{n-1}}{2^{2n-1}}+\cdots +\frac{b_2}{2^{n+2}} \right) \right)}\left| 1 \right>}{\sqrt{2}}
\\& \,\,\,\,=\frac{\left| 0 \right> +e^{2\pi ia\cdot \frac{b_n}{2^1}}\left| 1 \right>}{\sqrt{2}}\otimes \frac{\left| 0 \right> +e^{2\pi ia\cdot \left( \frac{b_n+2b_{n-1}}{2^2} \right)}\left| 1 \right>}{\sqrt{2}}
\otimes \cdots \otimes
\\& \qquad \frac{\left| 0 \right> +e^{2\pi ia\cdot \left( \frac{b_n+2b_{n-1}+\cdots +2^{n-1}b_1}{2^n} \right)}\left| 1 \right>}{\sqrt{2}}\otimes \cdots \otimes 
\\& \qquad \frac{\left| 0 \right> +e^{2\pi ia\cdot \left( \frac{b_n+2b_{n-1}+\cdots +2^{n-1}b_1}{2^{2n}} \right)}\left| 1 \right>}{\sqrt{2}}
\\& \xrightarrow{t_n}\frac{\left| 0 \right> +e^{2\pi ia\cdot \frac{b_n+2b_{n-1}+\cdots +2^{n-1}b_1}{2^1}}\left| 1 \right>}{\sqrt{2}} \otimes
\frac{\left| 0 \right> +e^{2\pi ia\cdot \left( \frac{b_n+2b_{n-1}+\cdots +2^{n-1}b_1}{2^2} \right)}\left| 1 \right>}{\sqrt{2}}
\\& \ \qquad \otimes \cdots \otimes 
\frac{\left| 0 \right> +e^{2\pi ia\cdot \left( \frac{b_n+2b_{n-1}+\cdots +2^{n-1}b_1}{2^n} \right)}\left| 1 \right>}{\sqrt{2}}
\otimes \cdots \otimes
\\& \ \qquad \frac{\left| 0 \right> +e^{2\pi ia\cdot \left( \frac{b_n+2b_{n-1}+\cdots +2^{n-1}b_1}{2^{2n}} \right)}\left| 1 \right>}{\sqrt{2}}
\\& \ \ =\bigotimes_{l=1}^{2n}{\frac{\left| 0 \right> +e^{\frac{2\pi ia\cdot b}{2^l}}\left| 1 \right>}{\sqrt{2}}}
\\& \ \ \xrightarrow{IQFT}a\cdot b
\end{aligned}
\end{equation}

\section{Results}
Classical matrix multiplication logic proceeds by row–column multiplication. For the product matrix \(C = A \cdot B\), each element \(c_{ij}\) is defined as the inner product of the \(i\)th row of matrix \(A\) and the \(j\)th column of matrix \(B\):
\begin{equation}
c_{ij}=\sum_k^{}{a_{ik}b_{kj}}
\label{eq-matrixmul}
\end{equation}
In the framework proposed here, this vector dot product is implemented using adder and multiplier circuits customized on the basis of the QFT.

By utilizing parameterized rotation gates, our adder directly applies phase operations to data, bypassing the unnecessary register extraction found in standard QFT-based designs. Consequently, the proposed adder not only streamlines the qubit count from $2n+1$ to $n+1$ but also achieves a linear gate complexity of $O(n)$, offering a decisive advantage over the $5n-3$ gates in classical adders. Regarding the multiplier, we integrate classical column-wise logic with an optimized control scheme that converts two-bit control gates into single-bit versions. This refinement reduces the qubit footprint from $4n$ to $3n$ and brings the gate complexity down from $O(n^3)$ to $O(n^2)$, markedly more efficient than the classical $6n^2$ benchmark. As detailed in Table 1, our approach exploits the inherent parallel capabilities of parameterized gates, significantly reducing hardware resource demands and paving the way for practical, large-scale quantum matrix multiplication.

\begin{figure*}
	\centering
	\includegraphics[width=1.0\textwidth,height=0.5\textwidth]{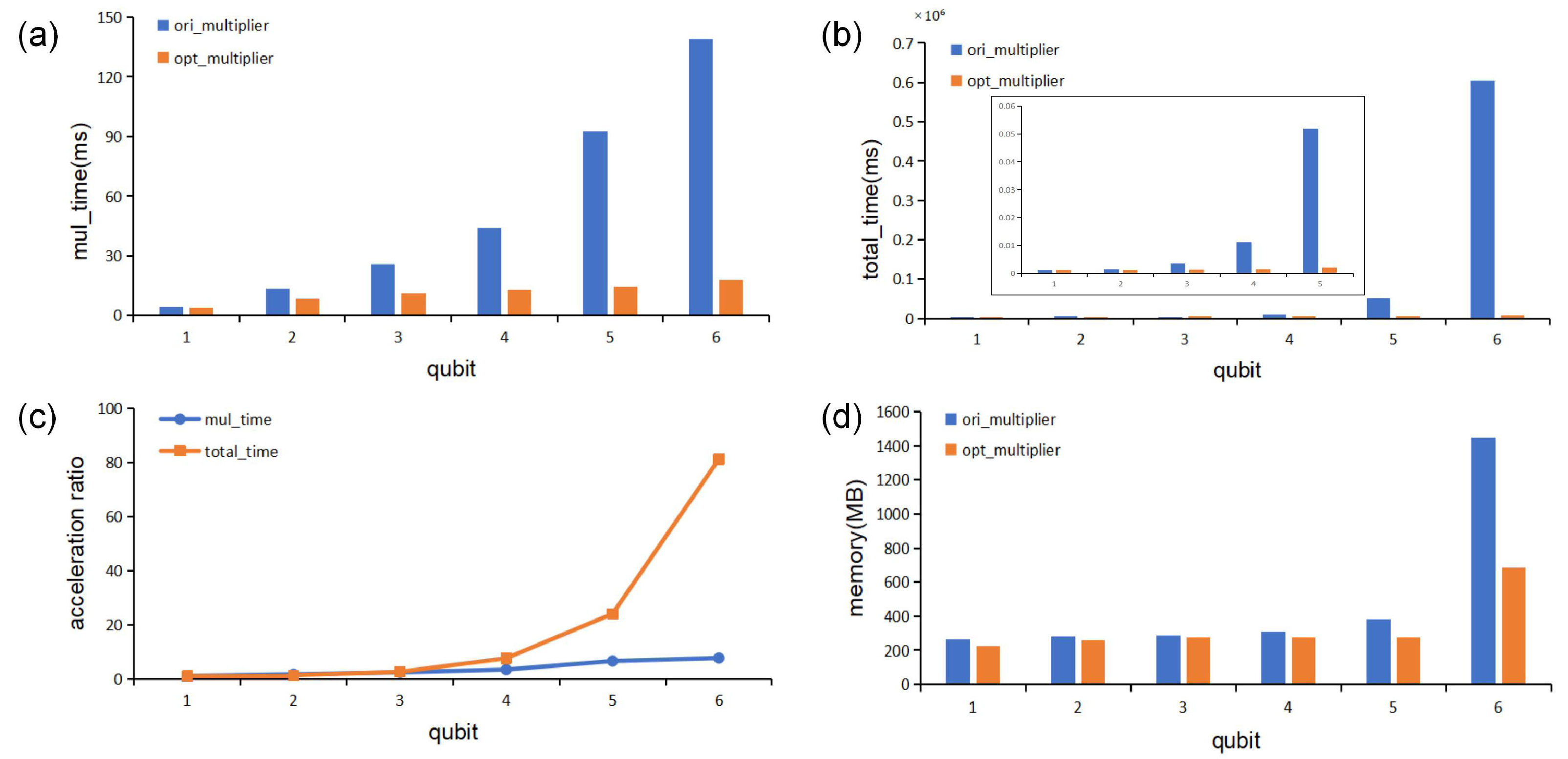}
	\caption{Performance comparison of quantum matrix multiplication with optimized multiplier; (a) Comparison of multiplication time consumption; (b) Comparison of total time consumption; (c) The time acceleration ratio; (b) Comparison of memory consumption.}
	\label{fig:Fig4_mul_speedup}
\end{figure*}

To validate the feasibility of the proposed quantum matrix multiplication, we performed circuit experiments using matrices with non‑negative integer entries. Taking matrices \(A = \begin{bmatrix} 1 & 2 \\ 3 & 4 \end{bmatrix}\) and \(B = \begin{bmatrix} 2 & 3 \\ 4 & 5 \end{bmatrix}\) as an example, we constructed the corresponding algorithmic circuits on the IBM quantum simulator and the PennyLane platform. The experimental measurement results are shown in Fig. \ref{fig:Fig2_result}, with data presented in hexadecimal format and its reverse order. For instance, the first measurement result, \(0x500\), corresponds to the binary sequence 010100000000; after reversing the sequence, this yields the decimal value 10, thereby confirming the accuracy of the computation.

\begin{figure*}
	\centering
	\includegraphics[width=1.0\textwidth,height=0.5\textwidth]{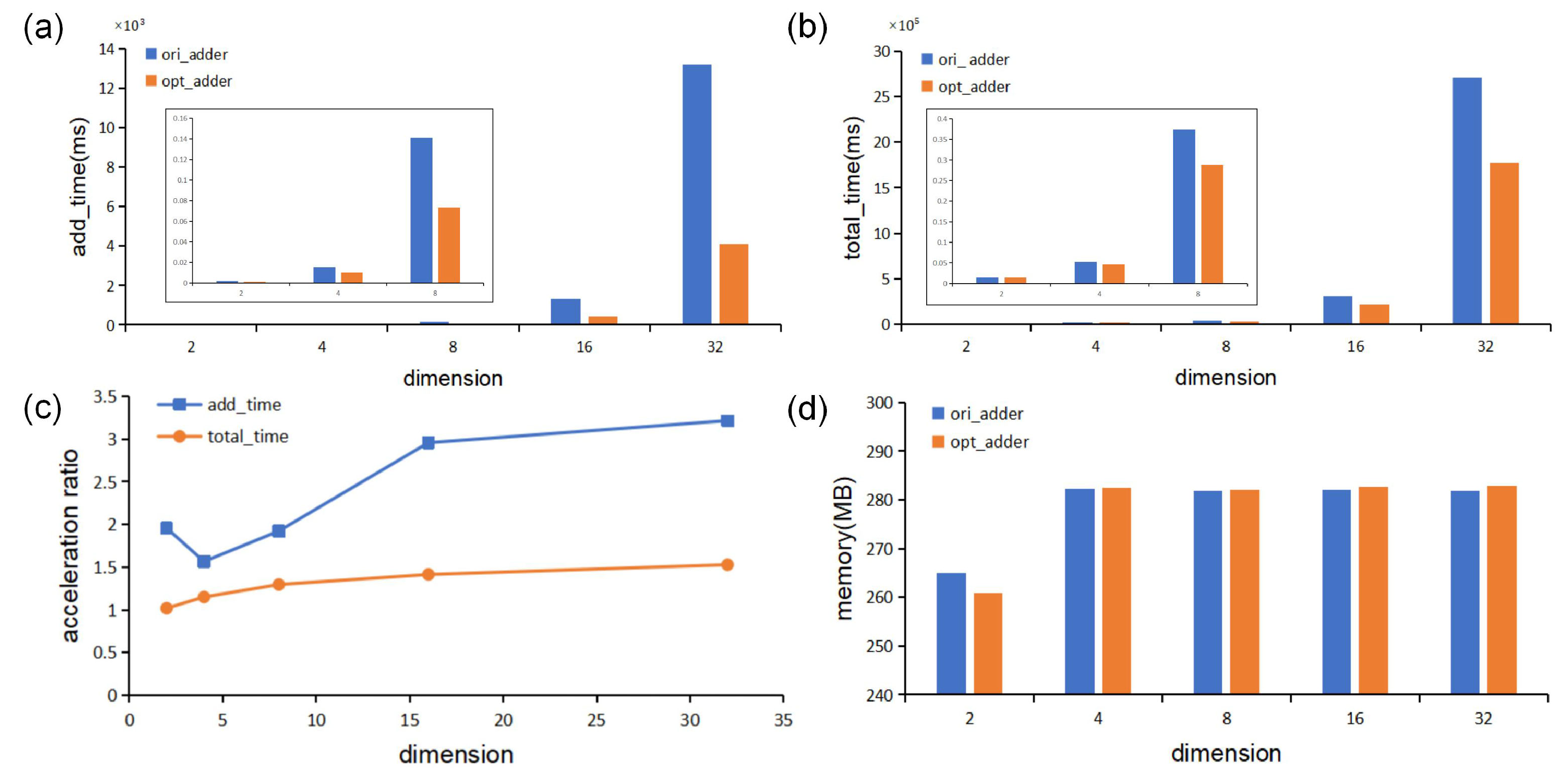}
	\caption{Performance comparison of quantum matrix multiplication with optimized adder; (a) Comparison of addition time consumption; (b) Comparison of total time consumption; (c) Line chart of time acceleration ratio; (b) Comparison of memory consumption.}
	\label{fig:Fig3_add_speedup}
\end{figure*}

To validate the acceleration performance of the optimized quantum adder in iterative operations, we performed scaling tests across increasing matrix dimensions. With matrix elements fixed at 7 (3-qubit representation), we compared an adder-only optimized version against a fully optimized (adder and multiplier) implementation. Analysis of Fig. \ref{fig:Fig3_add_speedup}(a) and \ref{fig:Fig3_add_speedup}(b) reveals that the performance gains of our scheme become more pronounced at higher dimensions, particularly regarding addition and total runtime. This improvement is quantified by the speedup curve in Fig. \ref{fig:Fig3_add_speedup}(c). Notably, Fig. \ref{fig:Fig3_add_speedup}(d) indicates that these computational gains are achieved without increasing memory overhead, as the optimized algorithm maintains resource parity with the baseline scheme.

Furthermore, as the number of qubits occupied by the multiplier increases, the advantages of the optimized quantum multiplier become increasingly apparent. We therefore fixed the experiment to \(2 \times 2\) matrix multiplication and systematically explored the relationship between bit width and acceleration performance by varying the multiplier from 1 to 32 (corresponding to \(n = 1, 2, \dots, 6\) qubits). Comparative data in Fig. \ref{fig:Fig4_mul_speedup}(a) and \ref{fig:Fig4_mul_speedup}(b) clearly show that, compared with the original multiplier, the optimized scheme achieves substantial reductions in both multiplication time and total execution time, with these time savings becoming more pronounced as the number of bits increases. The speedup curve in Fig. \ref{fig:Fig4_mul_speedup}(c) further confirms that the speedup ratio for total execution time rises significantly with problem scale, peaking at 6 bits and demonstrating excellent computational efficiency. Meanwhile, the memory monitoring record in Fig. \ref{fig:Fig4_mul_speedup}(d) indicates that the optimized multiplier consumes far fewer memory resources than the original version when handling higher-bit operations, successfully achieving dual optimization of computational speed and space utilization.

\subsection{Quantum Strassen Matrix Multiplication Algorithm}
The QFT‑based matrix multiplication framework is not limited to the basic inner‑product method but can be extended to other classical matrix multiplication algorithms, such as the Strassen algorithm \cite{strassen1969gaussian}. Here we implement a quantum version of the Strassen algorithm by integrating the optimized quantum adder and multiplier. To address the issue of negative intermediate values arising in this algorithm, we improve the addition logic by reserving the most significant bit as a sign bit and introducing additional compensation operations, thereby enabling quantum circuits to handle both positive and negative numbers.

\begin{figure*}[t]
	\centering
	\includegraphics[width=1.0\textwidth]{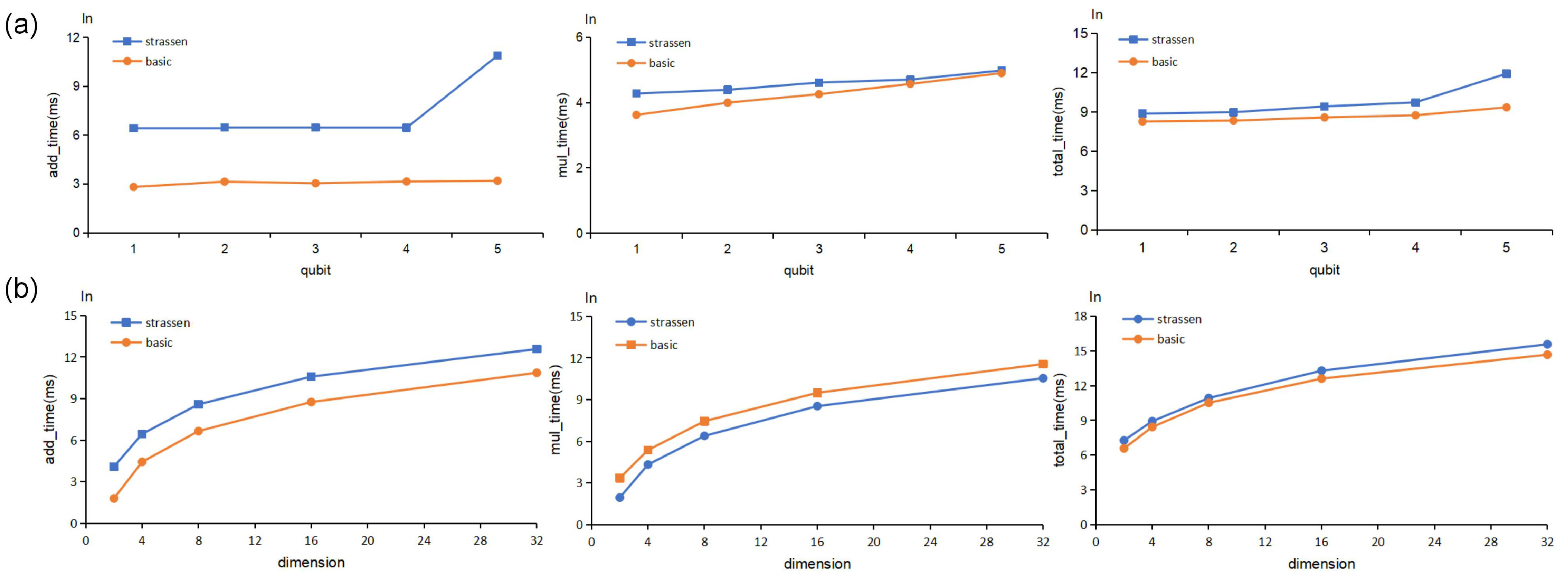}
	\caption{Time consumption comparison between the basic universal quantum matrix multiplication and quantum Strassen matrix multiplication; (a) Comparison of addition, multiplication, and total time consumption for $4\times4$ matrix multiplication; (b) Comparison of addition, multiplication, and total time consumption as the dimension of input matrices increases.}
	\label{fig:Fig5_strassen}
\end{figure*}

Following the basic principles of the Strassen algorithm, its quantum implementation inevitably generates a large number of intermediate variables during computation, imposing higher demands on qubit storage and arithmetic capabilities. To quantify its performance characteristics, we conducted two sets of comparative experiments (see Fig. \ref{fig:Fig5_strassen}). In the first experiment, we fixed the matrix dimension to \(4 \times 4\) and performed matrix multiplications with elements equal to 1, 2, 4, 8 and 16 (corresponding to qubit counts \(n = 1, 2, \dots, 5\)). We compared the differences in addition, multiplication and total time consumption between the baseline general‑purpose quantum matrix multiplication and the quantum Strassen matrix multiplication (Fig. \ref{fig:Fig5_strassen}(a)). In the second experiment, we set every element of both matrices \(A\) and \(B\) to 1 and observed the differences in addition, multiplication and total time consumption between the two algorithms as the matrix dimension increased (Fig. \ref{fig:Fig5_strassen}(b)).

Analysis of the experimental data in Fig. \ref{fig:Fig5_strassen} shows that although the quantum Strassen algorithm effectively reduces execution time in the multiplication phase, the recursive computation process generates a large number of intermediate variables, which directly leads to a sharp increase in the computational load of addition and a significant rise in the demand for qubit storage resources. In the \(4 \times 4\) comparative experiment, as the element bit width increased from 1 to 5 bits, the Strassen algorithm exhibited faster growth in both addition overhead and total time consumption than the baseline general‑purpose quantum matrix multiplication, despite reducing multiplication time. Similarly, in the dimension‑scaling experiment, the complexity of the addition logic increased in parallel with the matrix size. These experimental results indicate that the quantum Strassen algorithm trades increased addition resource overhead for improved multiplication efficiency. In the current context of limited quantum resources, the choice of algorithm must therefore be carefully evaluated based on hardware constraints such as qubit count and gate depth.

\subsection{Comparison of Fidelity in Matrix Multiplication under Noise Simulation}
Although experimental data from ideal simulators have validated the efficiency of the optimized algorithms, noise remains the primary barrier to practical deployment in the current era before fault‑tolerant quantum computing matures. Accordingly, in this chapter we introduce a simulation environment that incorporates realistic physical noise characteristics to comprehensively evaluate the noise resilience of the proposed universal quantum matrix multiplication framework.

Our noise model is constructed entirely upon the physical parameters of operational IBM quantum hardware, governed by two primary metrics. First, the decoherence times—energy relaxation ($T_1$) and phase dephasing ($T_2$)—are parameterized by applying a dynamic scaling factor in the range $[0.1, 1.0]$ to the hardware baselines ($\text{mean } T_1 = 233\ \mu\text{s}, \text{mean } T_2 = 160\ \mu\text{s}$), representing high-to-low noise regimes. Under independent scaling, the system dynamically enforces the physical boundary $T_2 \le 2T_1$ to exclude non-physical states, ensuring realistic decoherence dynamics. Second, the single-qubit and two-qubit gate fidelities are swept within $[0.9800, 0.9999]$. This range comprehensively captures various hardware developmental phases, transitioning from noisy intermediate-scale devices to high-fidelity thresholds nearing fault tolerance.

\begin{figure*}
	\centering
	\includegraphics[width=1.0\textwidth]{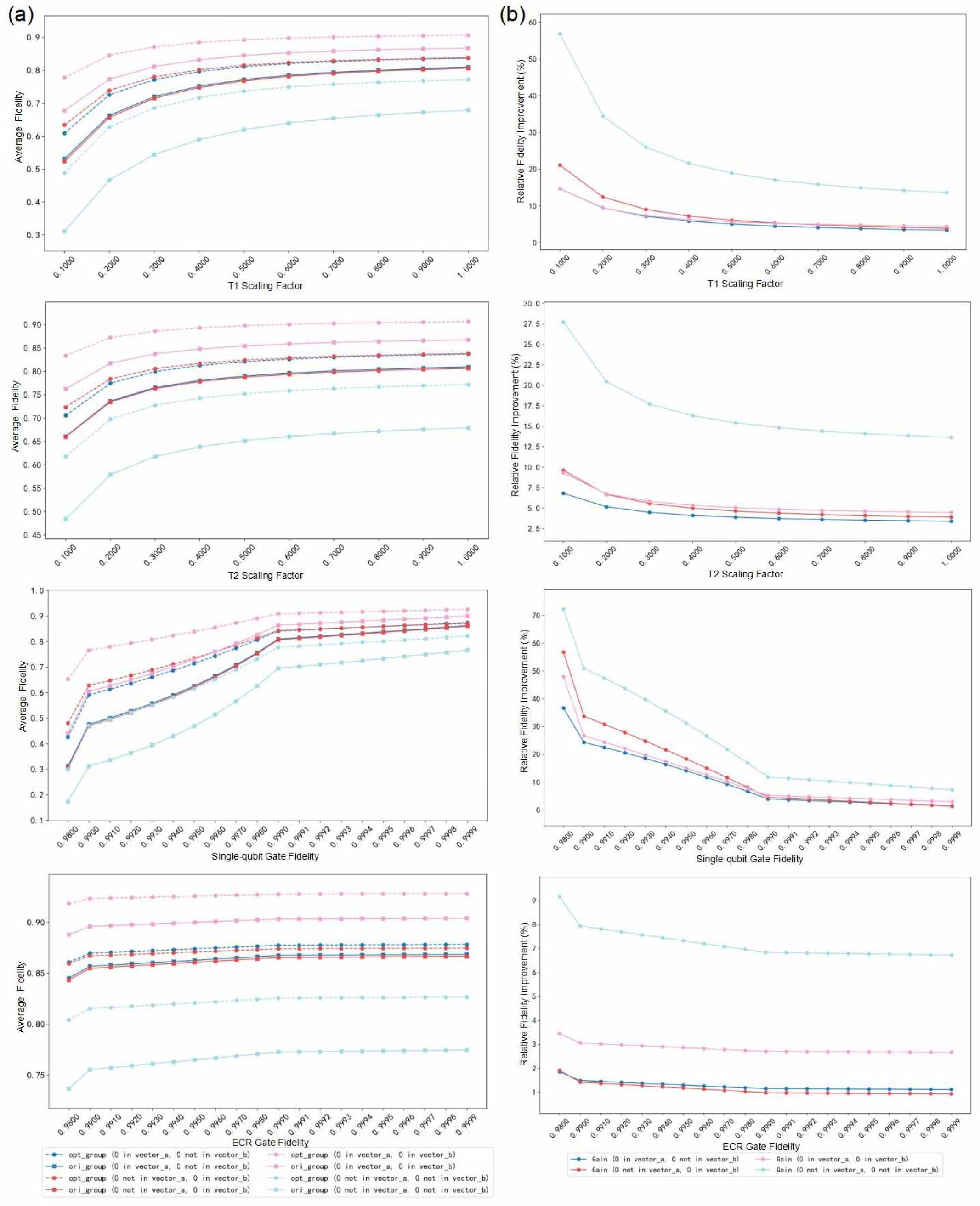}
	\caption{Evaluation of noise resilience for universal quantum matrix multiplication under various hardware noise regimes; (a) Performance curves of average fidelity ($F$) across varying $T_1$, $T_2$ decoherence times and gate fidelities; (b)Performance curves of relative fidelity growth rate varying $T_1$, $T_2$ decoherence times and gate fidelities.}
	\label{fig:Fig6_Noise_Simulation}
\end{figure*}

To systematically evaluate the performance of the optimized adder and multiplier, we introduced two core metrics—average fidelity (Fig. \ref{fig:Fig6_Noise_Simulation}(a)) and percentage fidelity gain (Fig. \ref{fig:Fig6_Noise_Simulation}(b))—to conduct a multi‑dimensional comparison of noise resilience between the scheme integrating the optimized adder and multiplier (opt-group) and the scheme based on the original adder and multiplier (ori-group). By finely controlling the evolution ranges of $T_1$, $T_2$ and logic gate fidelities, we systematically simulated the hardware progression from the degraded state of early high‑noise hardware to the near‑ideal fault‑tolerant threshold, thereby comprehensively quantifying the performance boundaries and robustness advantages of the optimized architecture under various physical noise barriers.

The experimental dataset covers all permutation inner products in three dimensions (values 0–3), comprising a total of 2,048 vector inner‑product pairs. The average fidelity of the final quantum circuit output is evaluated by measuring the overlap between the noisy density matrix \(\rho_{\text{noise}}\) and the ideal noise‑free density matrix \(\rho_{\text{ideal}}\), according to the formula:
$$F(\rho_{\text{ideal}}, \rho_{\text{noise}}) = \left( \text{Tr} \sqrt{\sqrt{\rho_{\text{ideal}}} \rho_{\text{noise}} \sqrt{\rho_{\text{ideal}}}} \right)^2$$
Second, the relative fidelity improvement is used to directly quantify the relative improvement in noise resilience of the opt-group over the ori-group. It is defined as:
$$\text{Relative Fidelity Improvement
 }(\%) = \frac{F_{\text{opt}} - F_{\text{ori}}}{F_{\text{ori}}} \times 100\%$$
where \(F_{\text{opt}}\) and \(F_{\text{ori}}\) represent the average fidelities measured for the opt-group and the ori-group, respectively, under identical noise configurations.

Analysis of the fidelity results of Fig. \ref{fig:Fig6_Noise_Simulation}(a) shows that, regardless of whether the independent variable is the scaling factor of \(T_1\) and \(T_2\) (from 0.1 to 1.0) or the fidelities of single‑qubit gates and ECR gates (from 0.9800 to 0.9999), the average fidelity of both the opt-group and the ori-group exhibits a steady increase and gradual convergence as hardware parameters improve. In the extreme high‑noise regime characterized by low scaling factors or low gate fidelities (for example, \(T_1 = 0.1\) or single‑qubit gate fidelity of 0.9800), although the absolute fidelities of both groups are at their lowest owing to decoherence and operational errors, the opt-group (dashed line) consistently maintains a significantly higher absolute fidelity than the ori-group (solid line), with a flatter curve slope, demonstrating superior resilience against hardware noise.

Analysis of the relative fidelity improvement results of Fig. \ref{fig:Fig6_Noise_Simulation}(b) further reveals the sensitivity of the optimized framework to different noise characteristics. Across the full parameter evolution range, as noise parameters gradually improve, the relative fidelity improvement consistently shows a pattern of peaking in the high‑noise regime and then declining and converging as the noise decreases. At the most extreme noise points, where the physical hardware environment is poorest, the redundant circuits in the original scheme (ori-group) severely amplify the accumulation of errors, causing its baseline fidelity to collapse substantially. As hardware parameters progressively improve, the absolute difference between the two schemes narrows, but the opt-group still retains a measurable advantage. This overall trend of the relative fidelity improvement curve evolving from steep to smooth reflects the universal optimization benefit of our proposed algorithmic framework across different physical noise mechanisms.

We observed that the inclusion of zero elements triggers pronounced shifts in the fidelity of matrix multiplication operations. Across all noise regimes in Fig. \ref{fig:Fig6_Noise_Simulation}, groups with zero elements consistently outperform non-zero groups in fidelity. This marked difference is rooted in the dynamic scaling of effective circuit depth under varying data profiles. In cascaded quantum arithmetic architectures, zero-valued inputs cause the control qubits of many multi-bit gates to remain inactive, thereby preventing the execution of the associated phase rotations. This inherent bypassing mechanism partially conceals the structural redundancies of the baseline design, allowing the conventional scheme to maintain an inflated baseline fidelity under noise. Therefore, when dealing with sparse matrices, the high density of zero elements effectively shields the redundant logic of conventional circuits via this control-bit lockouts, ultimately compressing the absolute fidelity differential between the two architectures.

\section{Discussion}


Matrix multiplication is a foundational algorithm in numerical computation, linear algebra, and even scientific computing. Due to its intrinsic high parallelism, the development of quantum-enhanced matrix multiplication algorithms under the quantum computing framework, capable of surpassing classical matrix multiplication methods, holds conceivable broad prospects and immense potential. This is our first step toward an even more efficient universal quantum matrix multiplication, demonstrating that the concept is feasible and worthy of continuous exploration and optimization. Clearly, future work will focus on investigating the potential impact of quantum matrix multiplication, including the potential for acceleration or reduction in hardware resources required for machine learning model training, particularly in the current era of Noisy Intermediate-Scale Quantum (NISQ) computers. Therefore, at this historical stage, it is valuable to further explore the quantum-classical hybrid architecture for matrix multiplication. We anticipate that practical implementations of quantum-classical hybrid matrix multiplication will first be realized on NISQ machines in the near future, followed by continuous optimization to adapt to computationally intensive scenarios such as large-scale model training. Following these advancements, we optimistically project that breakthroughs in quantum error correction will progressively unlock the formidable capabilities of fault-tolerant quantum computing. Consequently, quantum matrix multiplication algorithms that demonstrate verifiable quantum speedup are poised to achieve substantial reductions in computational costs within industrial applications while ushering in a transformative computational revolution throughout the AI ecosystem, encompassing pattern recognition and beyond.


\section*{Acknowledgment}

This work was supported by Tianjin Natural Science Foundation of China (20JCYBJC00500) and the Science \& Technology Development Fund of Tianjin Education Commission for Higher Education (2018KJ217).

\section*{Code availability}
The code was built using the Pennylane \cite{bergholm2018pennylane} platform. The code is available at \url{https://github.com/QML-TGU/QMM} \cite{matrix_multiplication_code}.



\bibliography{bibliography}
\end{document}